# One Step at a Time: Does Gradualism Build Coordination?[*]

Maoliang Ye[†], Jie Zheng[‡], Plamen Nikolov[§], Sam Asher[**]

[*] The authors thank the editors Axel Ockenfels and Uri Gneezy, an anonymous associate editor, and three anonymous referees for helpful comments, which significantly improved the paper; Alberto Alesina, Jim Alt, Nejat Anbarci, Abhijit Banerjee, Max Bazerman, Iris Bohnet, Hannah Bowles, Gary Charness, Raj Chetty, Soo Hong Chew, Kim Sau Chung, Lu Dong, Ninghua Du, Dirk Engelmann, Paolo Epifani, Liangcong Fan, Ernst Fehr, Qiang Fu, Daniel Friedman, John Friedman, Roland Fryer, Francis Fukuyama, Sen Geng, Francesca Gino, Edward Glaeser, Brett Graham, Xiangting Hu, Wei Huang, Torben Iversen, Garett Jones, Jianpei Li, Zhi Li, Pinghan Liang, Jingfeng Lu, Yuichiro Kamada, Larry Katz, Rongzhu Ke, Judd Kessler, David Laibson, Jeffrey Liebman, Jaimie Lien, Tracy Xiao Liu, Erzo F. P. Luttmer, Brigitte Madrian, Juanjuan Meng, Louis Putterman, Alvin Roth, Jason Shachat, Kenneth Shepsle, Andrei Shleifer, Monica Singhal, Qianfeng Tang, Dustin Tingley, Ruixin Wang, YunWang, Roberto Weber, Lijia Wei, Xi Weng, Chunlei Yang, Lan Yao, Tristan Zajonc, Richard Zeckhauser, Yao Zeng, and Hongliang Zhang; and participants at various meetings and seminars at Harvard University, Harvard Business School, Massachusetts Institute of Technology, Peking University, Tsinghua University, Fudan University, Shanghai University of Finance and Economics, Xiamen University, Shanghai Jiaotong University, Zhejiang University, Southwestern University of Finance and Economics, University of International Business and Economics, Southeastern University, Sun Yat-sen University, University of Nottingham Ningbo China, Hong Kong Baptist University, and Renmin University of China for valuable comments at different stages of this project. The authors also thank Lorenzo Casaburi for input in the early stage of our study; the School of Economics at Renmin University of China, and School of Economics and Wang Yanan Institute for Studies in Economics at Xiamen University for providing the laboratory; and Matthew Bonci, Dayne Feehan, Xuelin Li, and Yuqi Zheng for providing excellent research assistance. This paper was accepted when Sam Asher was at the Development Research Group, World Bank.
[†] Corresponding author. Department of Public Finance, School of Economics, Wang Yanan Institute for Studies in Economics, MOE Key Laboratory of Econometrics, and Fujian Key Laboratory of Statistical Science, Xiamen University, Xiamen 361005, Fujian, China. email: maoliang.ye@post.harvard.edu. Tel: (+86) 592-2188827.
[‡] Department of Economics, School of Economics and Management, Tsinghua University, Beijing, China 100084. email: jie.academic@gmail.com.
[§] Department of Economics, State University of New York, Binghamton, NY 13902. email: pnikolov@post.harvard.edu. IZA Institute of Labor Economics.
[**] World Bank, 1818 H Street, NW, Washington, DC 20433, email: sasher@worldbank.org.

# One Step at a Time:
# Does Gradualism Build Coordination?

**Abstract**

This study investigates a potential mechanism to promote coordination. With theoretical guidance using a belief-based learning model, we conduct a multi-period, binary-choice, and weakest-link laboratory coordination experiment to study the effect of gradualism – increasing the required levels (stakes) of contributions slowly over time rather than requiring a high level of contribution immediately – on group coordination performance. We randomly assign subjects to three treatments: starting and continuing at a high stake, starting at a low stake but jumping to a high stake after a few periods, and starting at a low stake while gradually increasing the stakes over time (the Gradualism treatment). We find that relative to the other two treatments, groups coordinate most successfully at high stakes in the Gradualism treatment. We also find evidence that supports the belief-based learning model. These findings point to a simple mechanism for promoting successful voluntary coordination.

# 1. Introduction

Coordination is at the core of a wide variety of economic activities and organizational performance (Schelling, 1960; Arrow, 1974).[1] However, efficient coordination outcomes are often difficult to attain without effective coordination mechanisms (Van Huyck, Battalio & Beil, 1990; Knez & Camerer, 1994; Cachon & Camerer, 1996; Weber, 2006; Devetag & Ortmann, 2007). Managers routinely face the challenge of how best to induce a high level of effort and facilitate coordination across multiple agents both within and across organizations, which has been a central issue in the practice and science of organizational management (Van De Ven, Delbecq & Koenig, 1976; Lounamaa & March, 1987; Smith, Carroll & Ashford, 1995; Siemsen, Balasubramanian & Roth, 2001; Rico et al., 2008; Okhuysen & Bechky, 2009). Leading business magazines such as the Harvard Business Review routinely run articles on how to promote teamwork and coordination (e.g., Hackman, 2009; Prusak, 2011).

In this paper we propose and test one mechanism to promote coordination, which we call gradualism: starting small and gradually increasing the stake of a coordination project within a fixed group. The corresponding hypothesis is that allowing agents to coordinate first on small and easy-to-achieve goals (projects) and slowly increasing the level of goals, facilitates subsequent coordination on otherwise hard-to-achieve outcomes.

Gradualism is employed by managers and leaders in a wide range of real-world settings. Team building often adopts a gradual method: to help build coordination in collaborative projects involving pivotal efforts from all members, new employees and teams are initially given smaller or easier tasks which ensure that they can then coordinate well in larger or harder tasks later. For example, law enforcement units, police Special Weapons and Tactics (SWAT) teams, as well as special military groups, are initially given ordinary low-stake tasks until proven effective before they are assigned to resolve more critical high-stake situations. In entrepreneurship, the success of a startup venture requires the effort and input from all partners with various expertise and resources, and the growth path involves starting very small (e.g., little capital), and then progressing through funding rounds where larger and larger investments are made (Zhou & Su, 2011). Microfinance institutions often offer group loans that start small and increase in size upon successful repayment by all members of the group (Armendariz & Morduch, 2010). Last, leaders of international organizations also adopt the gradualist approach to facilitate international coordination. Abbott and Snidal (2002) highlights the function of the gradualist approach in the development of the 1997 OECD Anti Bribery Convention. Combating bribery of foreign public officials in international business transactions requires inputs from all countries involved; otherwise, the corrupted business can easily circumvent the campaign and exercise in the countries with least effort on anti-corruption. Thus, the international anti-bribery collaboration reflects a multi-player coordination problem, which involves the strategic uncertainty about the effort level other partner countries are willing to exert. Breaking the final goal into a series of steps has allowed the countries to avoid the failure in previous big-bang approaches which sought an immediate international treaty requiring others states to adopt the equivalent anti-bribery regulation as in the US. International agreements on law enforcement, trade, arms reduction, and environmental coordination have also adopted a similar gradualist approach (Benyon, 1994; Langlois & Langlois, 2001).

---

[1] In this study, we focus on those coordination games with Pareto-ranked equilibria.

We conduct a computer-based laboratory coordination experiment with repeated interactions. In each period, every subject is endowed with a certain amount of points (monetary units in the laboratory) and is asked to participate in a group project with a certain stake level. Each subject has two options: (1) to contribute the exact pre-set amount (i.e., the stake) or (2) to contribute nothing. In each period, each subject realizes an extra return only when all group members contribute to the project; otherwise, each member ends up with the initial endowment minus his/her contribution. This set-up is generally referred to as the minimum-effort or weakest-link coordination game: the payoff depends on the effort of the individual and the minimum effort of all group members. The stage game in each period is a multi-player stag hunt because of the binary choice feature available to each player.[2]

We assign the subjects to three main treatments of differing stake patterns. In particular, the stake patterns differ in the first six periods but feature an identical high stake for the next six periods. The first treatment, which we call Big Bang, features a constant high stake for all 12 periods. The second treatment, called Semi-Gradualism, features a constant low stake for the first six periods and then a high stake for the next six periods. In our third and key treatment, termed Gradualism, we increase the stake in each of the first six periods in small amounts until the highest stake is reached in Period 7. We exploit this design to address how the pattern of varying stake levels influences group coordination at high-stake levels. Specifically, we test the effect of gradualism on coordination performance in the high-stake periods. See Figure 1 for a graphical illustration of all treatments.

[Figure 1 about here]

We propose a belief-based learning model which predicts that gradualism attains more successful coordination at a high-stake level than alternative treatments, which is confirmed by our experimental results. In terms of magnitude, the effects are large: 61.1% of the Gradualism groups successfully coordinate in the final period, whereas only 16.7% and 33.3% of Big Bang and Semi-Gradualism groups do so, respectively. The Semi-Gradualism treatment fails to foster high-stake coordination compared with the Gradualism treatment. Our findings suggest that for a group to establish successful coordination at a high-stake level, it is better to begin at a low-stake level and, equally important, to increase the stake level slowly.

---

[2] Bolton, Feldhaus, and Ockenfels (2006) provide a brief summary of the stag hunt literature.

When a person interacts with others in a group, he/she not only learns about group members but may also develop beliefs about the contributing tendencies of an average person from the general population. Thus, we introduce a second stage to the experiment. The subjects from various treatments in the first stage were randomly reshuffled into new groups when they entered the second stage of the game, and then continued to play at the same stake level as the end of the first stage. We hypothesize that subjects who were in the gradualism treatment may contribute more when they enter the new groups because they have higher beliefs about others' contribution possibilities. Subjects may also adapt their behavior over time as their beliefs update in the second stage.

We find that subjects in the Gradualism treatment, who are more likely to experience successful coordination at the end of the first stage than subjects in the other treatments, are 12.2 percentage points more likely to contribute upon entering a new group when we reshuffle subjects from all treatments into new groups in the second stage. However, subjects who were initially in the Gradualism treatment become less likely to contribute when they find that their contributions are not rewarded in the new environment, possibly because the new group members may have had different coordination outcomes previously. This result provides suggestive evidence of the role of beliefs which players carry from old groups to new ones and later update based on the performance of the new groups.

To further explore the potential channel through which gradualism fosters high-stake coordination, we conduct a supplementary experiment akin to the main experiment except that we explicitly elicit the beliefs of the subjects regarding the probability that other group members contribute in each period. The results are consistent with the belief-based learning model: subjects form their initial beliefs based on the initial stake, make their contribution decisions based on their beliefs, and update their beliefs after observing the coordination outcome in each period.

The present study is the first that clearly tests the role of exogenous gradualism in coordination within a fixed group. The present study adopts an exogenous setting of stake paths and explores whether gradualism works better rather than whether players choose gradualism. By randomizing subjects into various treatments, we avoid the problem of self-selection when gradualism is endogenously chosen. We find that the coordination success rate is indeed higher under the gradualism mechanism, which may help explain why gradualism is a popular practice in many real world settings.

The present study has two key contributions to the literature. First, our study contributes to the literature on coordination mechanisms. Economists have addressed ways to promote successful coordination via various mechanisms, such as communication (Cooper et al., 1992; Charness, 2000; Weber et al., 2001; Duffy & Feltovich, 2002, 2006), teams (Feri, Irlenbusch & Sutter, 2010), between-group competitions (Bornstein, Gneezy & Nagel, 2002; Riechmann & Weimann, 2008), voluntary group formation (Yang et al., 2017), voluntary reward (Yang et al., 2018), gradual organization growth (Weber, 2006), social identities (Chen & Chen, 2011; Chen, Li, Liu & Shih, 2014), information feedback (Berninghaus & Ehrhart, 2001; Devetag, 2003; Brandts & Cooper, 2006a), and transfer of learning across games (Devetag, 2005; Cason, Savikhin & Sheremeta, 2012). The present study examines how exogenous stake paths foster successful group coordination, thus proposing gradualism as an alternative mechanism. To the best of our knowledge, the only other studies that examine the effect of exogenous path dependence on group coordination are Weber (2006) and Romero (2015).[3] Second, our study contributes to the more general literature on dynamic simultaneous games with a gradualist feature, such as in the settings of public goods (Dorsey, 1992; Marx & Matthews, 2000; Kurzban et al., 2001; Duffy, Ochs & Vesterlund, 2007; Offerman & van der Veen, 2013; Oprea, Charness & Friedman, 2014) and prisoners' dilemmas (Andreoni & Samuelson, 2006). We provide detailed comparisons between the present study and the literature on dynamic simultaneous games in Section 2.

## 2. Relation with the Literature on Dynamic Simultaneous Games

To better understand our study and its contributions, we compare it to the theoretical and experimental literature on coordination games with varying paths and other dynamic simultaneous games (e.g., public goods games, prisoners' dilemma games) with a gradualist feature.

In a dynamic laboratory weakest-link coordination experiment, Weber (2006) studies the dynamics of organizational growth and finds that gradually expanding group size leads to more successful coordination compared to immediately starting with a large group. The present study differs from Weber (2006) in four major ways: (1) Weber studies the path of group size, whereas we vary the stake path and explore gradualism in coordination within a fixed group; (2) in our study each player's choice set in each period is binary; (3) a third treatment – Semi-Gradualism – is used, which explores whether a sudden stake increase negatively affects coordination; (4) our theoretical model is based on a belief-based learning framework, whereas Weber (2006) adopts a linear adaptive dynamics framework.

In another study, Romero (2015) finds that groups coordinate better when the cost is increasing to a specific level than when the cost is decreasing to that level. Our setting differs from his study in two ways: (1) we change the stake level, which indicates not only the cost but also the benefit, whereas the stake level is fixed in his study;[4] (2) we compare a slow increase with a sudden increase in stake as well as a high-stake

---

[3] Riedl, Rohde and Strobel (2011), Salmon and Weber (2011), and Yang et al. (2013) use what can be considered a general path dependence approach, but with endogenous paths. They all focus on the issue of group size. The exogenous stake path change in our design is novel.
[4] Other studies change either benefit (Brandts & Cooper, 2006b) or cost (Goeree & Holt, 2005), but not both. Our setup reflects those studies (Van Huyck et al., 1990; Knez & Camerer, 1994; Cachon & Camerer, 1996; Weber, 2006)

starting point, whereas in his study the speed of cost change is fixed. In addition, the theoretical models are different: we employ a belief-based learning model with incomplete information, whereas Romero (2015) assumes a quantal response equilibrium framework with bounded rationality.

Battalio, Samuelson, and Van Huyck (2001), in a laboratory setting, also explore stake size effects in coordination games. They focus on how various levels of an optimization premium (defined as the difference between the payoff of the best response and inferior response to an opponent's strategy) influence players' behavior differently, where the stake size is the optimization premium. In their study, the stake size is fixed within a treatment. By contrast, we allow the stake size, interpreted as the participation cost (as well as the net profit of success) of a stag-hunt coordination project, to vary over time within a treatment. In particular, we examine the role of gradually increasing the stakes in building large-stake coordination, which is not examined by Battalio, Samuelson, and Van Huyck (2001).

Within the literature on dynamic real time public goods provision, the gradualist approach in the present study differs from those of previous studies, in particular, Dorsey (1992), Marx and Matthews (2000), Kurzban et al. (2001), Duffy et al. (2007), and Oprea et al. (2014). In addition to our binary weakest-link structure, which markedly differs from these studies, another unique feature of our setup is that the public projects in our game are independent from one period to the next: contributions cannot accumulate over periods, and each project features its own target (stake). In the aforementioned studies on dynamic voluntary contribution to a single public project, players are allowed to contribute whenever and as much as they wish and to accumulate their contributions over the course of the project (before the end of the game, each period has no objective). Our study determines whether first working on smaller tasks facilitates the accomplishment of a large task in a given group, rather than whether dividing the call for contribution into multiple periods and allowing contributions to accumulate over time improves collective contributions. Although the aforementioned studies do relate to certain real-world examples (e.g., long-term fund drives), the present study is better aligned with the different, yet just as significant, real-world cases mentioned earlier. In those introductory real-world examples, the duration of the final high-stake project is relatively short and is not divisible into sub-periods to accumulate effort. Moreover, regular feedback about what other participants contribute to the final project is not provided. Instead, players face an independent project with a clear and smaller objective in each period other than the final high objective, and players assess how they performed on these small tasks after each period.

Offerman and van der Veen (2013) explore whether governmental subsidies geared toward promoting public good provisions should be abruptly introduced or gradually increased; in other words, given the

---

that fix the benefit-cost ratio 2:1 at equilibrium, but we simplify the decision choice to be binary and change the scale of the group project simultaneously through varying the stake. Yang et al. (2017) have a similar setup where the benefit-cost ratio is a constant 2:1 at equilibrium, but the scale increases with group size.

benefit of the public good, whether the individual cost of providing the public good should be decreased sharply or gradually. The results favor an immediate increase of subsidy: when the final subsidy level is substantial, the effect of a quick increase is stronger than that of a gradual increase. The present study differs from Offerman and van der Veen (2013) in three key ways. First, these authors focus on how the use of subsidies can stimulate cooperation after its initial failure. The mechanism used in the present study is distinct from their subsidy mechanism as it focuses on the variation of stake patterns instead of governmental subsidies. Second, our study manipulates the stake level that decides both the cost and the benefit of the public good, whereas in the setup of Offerman and van der Veen, only the cost changes. Third, the stake paths of the present study are non-decreasing, whereas paths of the other are non-increasing.

We outline the efficiency gains of gradualism more clearly than Andreoni and Samuelson (2006). Andreoni and Samuelson (2006) examine a twice-played prisoners' dilemma, in which the total stakes in two periods are fixed and the distribution of these stakes across periods can be varied. Both their theoretical and experimental results show that the best way is to "start small" with bigger stakes in the second period. However, cooperation is low for the period with a high stake in their experiment. One potential explanation of the advantage of our setup in promoting efficiency over that of Andreoni and Samuelson (2006) is that unlike their setup, our weakest-link structure does not allow free riding.

Several theoretical studies examine monotone games – multi-period games in which players are constrained to choose strategies that are non-decreasing over time (i.e., players must increase their respective contributions over time) (Gale, 1995, 2001; Lockwood & Thomas, 2002; Choi, Gale & Kariv, 2008). In contrast to these studies, our game employs a different feature – we require the stake level, rather than the contribution, to be non-decreasing.

Watson (1999, 2002) examines theoretically how "starting small and increasing interactions over time" is an equilibrium for dynamic cooperation with the option to break up unilaterally, which endogenously forces "starting small" in equilibrium. In contrast, through determining the stake path exogenously, we address whether gradualism promotes coordination at high-stake levels among fixed partners with no chance of breaking up (each player in a group can choose not to contribute in each period, but not to leave the group before the game ends), rather than whether players themselves choose to adopt a gradualist approach. In addition, our setup focuses on weakest-link coordination problems with no chance of free riding.

In summary, there are many organizational settings in which changing group sizes, group compositions, or other factors is difficult, but where a leader can start a group with an easier version of a task to provide reassurance and strengthen beliefs. This is a common practice and therefore worth investigating. We explore this mechanism experimentally together with a theoretical framework, and demonstrate that it can be a valuable tool for aiding high-stakes coordination.

## 3. Experimental Settings

### 3.1 Game Composition, Information Structure, and Payoff Structure

In July 2010, we conducted the laboratory experiment at the Renmin University of China in Beijing, China, with 256 subjects that were recruited through the bulletin board system and posters. The majority of the subjects were students from Renmin University and nearby universities.

The experiment consisted of 18 sessions, which were all computerized using the z-Tree software package (Fischbacher, 2007). Both the instructions (see S1 in the Supplemental Material) and the game information shown on the computer screen were in Chinese. In each session, we randomly assigned subjects to groups of four; our sample consisted of 64 groups in total.

The experiment included two stages: the first stage comprised twelve periods, whereas the second one comprised eight periods. Group members did not change within each stage, but subjects were randomly reshuffled into groups of four after the first stage; this rule was announced publicly in each session. The subjects were not told the exact number of periods in each stage. Instead, the subjects were told that the experiment would last from 30 minutes to one hour. This includes the time for sign-up, reading of instructions, completion of a quiz (designed to ensure that subjects understood the experimental rule), and final payment. Furthermore, at the beginning of each period, subjects knew the stake of the current period but not those of future periods. This condition replicates the circumstances of many real-world cases, in which people do not know the exact number of coordination opportunities or what is at stake in future interactions.

In each period, we endowed each subject with 20 points and asked each to give a certain number of points to the common pool of his/her assigned group. The required number of points could vary across periods, and subjects could only choose "to give the *exact* points required" (we use the natural term "give" rather than "contribute" in the instruction) which we refer to as *stake*, or "not to give" at all. If all members in a group contributed, then each member not only received the stake back, but also gained an extra return equal to the stake. If not all group members contributed, then each member finished the period with only his/her remaining points (i.e., the initial endowment in each period minus the contribution of the subject).

In summary, players earned points according to the following payoff function in each period, which is conditional on the actions of all players in the same group:

$$\pi_{i,t}(A_{i,t}, A_{-i,t}) = \begin{cases} 20 + S_t, & \text{if } A_{i,t} = C \text{ and } A_{j,t} = C, \forall j \neq i \\ 20, & \text{if } A_{i,t} = NC \\ 20 - S_t, & \text{if } A_{i,t} = C \text{ and } \exists j \neq i, \text{ s.t. } A_{j,t} = NC \end{cases}$$

where $\pi_{i,t}$ is $i$'s payoff at period $t$, $S_t$ is the stake at $t$. $A_{i,t}$ and $A_{j,t}$ are the actions of $i$ and $j$ at $t$, respectively ($i$ and $j$ are in the same group), and $A_{-i,t}$ denotes the list of actions of all players except $i$. $C$ represents "contribute," whereas $NC$ represents "not contribute." Thus, each period features a four-player stag hunt game, in which there exists one mixed-strategy Nash equilibrium, where each player contributes with probability $2^{-\frac{1}{3}}$, and two pure-strategy Nash equilibria: where all players contribute and where no player contributes. For all values of $S_t$, the secure equilibrium (all players choosing $NC$) is risk-dominant according to the Harsanyi-Selten definition. Except for the binary choice set in each period, our setup is consistent with the minimum effort coordination literature that has a 2:1 benefit–cost ratio at each pure-strategy equilibrium.

We did not allow communication across subjects. The preclusion of communication makes coordination among players difficult. The reason is that we are interested in studying how gradualism may help solve coordination difficulties in the absence of communication mechanisms.

At the end of each period, subjects knew whether all four group members (including himself/herself) contributed the stake for that period, but did not know the total number of group members who contributed (in case fewer than four members contributed). This design is consistent with the type of minimum-effort coordination games with limited information feedback, in which the only commonly available historical information to players is the minimum contribution of group members. This feature is popular in the contract theory literature, in which an imperfect observation of efforts is commonly assumed. By adopting this design, we can also increase the difficulty of coordination given other aspects of the experiment, and study whether gradualism can help overcome such a difficulty.

The final total payment to each player equaled the accumulated earnings over all periods plus a show-up fee. The exchange rate was 40 points per CNY 1. An average subject earned CNY 21–22 (around USD 3) including the show-up fee for the whole experiment, which covered ordinary meals for one to two days on campus. With regard to its purchasing power, the payment was comparable to those experiments conducted in other countries.

## 3.2 Treatment Group Assignments

Our experiment consisted of three main treatments: (1) Big Bang, (2) Semi-Gradualism, and (3) Gradualism. To isolate the wealth effect on the contribution of participants from the effect of the three main treatments in the second half (Periods 7–12) of the first stage, we also introduced a fourth High Show-up Fee treatment, which is identical to the Big Bang treatment except that we give subjects higher show-up fees. All groups in the three main treatments faced the same stake in the second half of the first stage, but stake paths differed for each treatment in the first half (Periods 1–6). We randomly assigned 12 subjects (three groups) into the three main treatments for eight of the 18 sessions. In the remaining 10 sessions, we randomly assigned 16 subjects (four groups) into the four treatments (three main treatments and the High Show-up Fee treatment). In total, we had 18, 18, 18, and 10 groups (or 72, 72, 72 and 40 subjects) in Big Bang, Semi-Gradualism, Gradualism, and High Show-up Fee treatments, respectively. Table S2 in the Supplementary Material shows that the randomization of treatment assignments worked well.

Figure 1 shows the game stakes across periods. For the Big Bang treatment, the stakes were always kept at the highest level, which was 14.[5] For the Semi-Gradualism treatment, we set the stakes at two for the first six periods and then we set them at the highest stake for the next six periods. Finally, for the Gradualism treatment, we gradually increased the stakes from 2 to 12 with a step of 2 for the first six periods, and we kept them fixed at the highest stake for the next six periods.

This experiment had a second stage, as mentioned earlier. The subjects from various treatments were randomly reshuffled into groups of four when they entered the second stage of the game. New group members did not necessarily come from the same group in the first stage; this rule was made common information. Within the second stage of the game, group compositions were fixed, and stakes were all set at the highest stake (14) for all periods and all groups.

At the start of the second stage, we notified each player that he/she would enter a new random group. At the end of each stage, we notified each player of the number of points he/she had accumulated to date.

At the end of the experiment, we asked subjects to complete a brief survey that collected information on age, gender, nationality, education level, concentration at school, working status, income, as well as their risk preferences over lotteries adopted from Holt and Laury (2002).[6]

Table 1 and Figure 1 illustrate the main experimental design.

[Table 1 about here]

## 4. Theoretical Predictions and Hypotheses
### 4.1 A Simple Belief-based Learning Model

The coordination problem in each period involves multiple equilibria. When multiple equilibria exist, the beliefs of the players are central in deciding which equilibrium outcome will be selected. Gradualism and alternative stake paths may matter in the coordination dynamics because they affect the beliefs of players: the stake level at the start of the game influences how players form their initial beliefs about the actions of others, whereas the stake path influences how players subsequently update these beliefs. Thus, we develop a simple belief-based learning model for theoretical predictions, although we do not rule out the possibility that other theoretical models may also be consistent with our experimental results (Levine & Zheng, 2015). The purpose of the model is to illustrate the intuition behind the efficacy of gradualism and its relation to beliefs, rather than providing a comprehensive model of belief-based choices in dynamic coordination games.

---

[5] We calibrated the highest stake level using 12 and 14, and finally opted for 12 in two sessions and 14 in 16 sessions. To make full use of the samples, we pooled all 18 sessions together in the analysis.

[6] We code the risk aversion attitude as the number of option A chosen in questionnaire by Holt and Laury (2002), ranging from 0 to 10. Thus, the larger the value, the more averse the subject is to risk. A simple regression of contribution decision on risk aversion shows the average marginal effect of risk aversion on contribution decision is -0.01, which suggests that if risk aversion increases from 0 (smallest possible value) to 10 (largest possible value), on average the contribution rate will fall by 10 percentage points. However, this effect is statistically insignificant when standard errors are clustered at either the group level (p=0.44) or at the subject level (p=0.45).

**Intuition**

Before we introduce the model and describe how the belief system works, we present the basic intuition as follows. In the belief-based learning framework, rational players have prior beliefs about the actions of others before the game starts, and update these beliefs based on the outcome of each period. The lower the stake level at the start of a game, the stronger are players' beliefs that others will contribute to the group project, and as a result, they are themselves more likely to contribute. Therefore, under the weakest-link payoff structure, the lower initial stakes will produce higher rates of contribution and success in group coordination (we define coordination success as the case that all group members contribute.) When groups successfully coordinate at a given stake level, players reinforce their beliefs about the likelihood that others will contribute at a stake level *less* than or equal to the current. Alternatively, coordination failure at a given stake level causes players to doubt that others will subsequently contribute at a stake level *greater* than or equal to the current level. Finally, when stakes increase in two consecutive periods, previous successful coordination at the lower stake level may (or may not) largely influence the posterior beliefs of the players regarding the actions of others at slightly (or substantially) higher stake levels. Thus, successful coordination at a low-stake level may not imply an immediate successful coordination at a high-stake level if the stake level increases dramatically. For this reason, slowly increasing the stake may better maintain coordination success.

The main aspects of the model are belief-based learning, myopia, and standard self-interested preference.[7] These assumptions allow us to focus on the belief updating process, which is an important feature in dynamic coordination games. A more general model of dynamic games is beyond the scope of the present study.

---

[7] We adopt the myopia (i.e., backward looking) assumption for three reasons. First, this assumption is often used in learning models, such as reinforcement learning (e.g., Roth & Erev, 1995), belief-based learning (e.g., Fudenberg & Levine, 1998), experience-weighted attraction learning (e.g, Camerer & Ho, 1998, 1999), and adaptive dynamics (e.g, Crawford, 1995; Weber, 2006). Second, myopia allows us to focus on the key process of belief updating. Third, in our experiment, players do not know the future stake and the number of periods, which may limit the potential of forward looking and more strategic play. Thus, the backward-looking assumption is a reasonable simplification.

**Setup**

The game structure and payoff rule in each period are as follows. There are $N$ periods. In each period, a group of $I$ risk-neutral players conduct a binary coordination task: each player can choose to contribute ($C$) or not to contribute ($NC$). The endowment per person in each period is $E$. The stake of the coordination task, $S_t \, (0 < S_t < E)$, may vary across periods. Thus, each player can choose to contribute either zero or exactly $S_t$ in each period, but no other amount of efforts to the group project. We adopt a minimum-effort (weakest-link, or more specifically, stag-hunt) payoff structure: the value of the project output for everyone is $\alpha S_t$ ($\alpha > 1$; we adopt $\alpha = 2$ in our experiment) if all $I$ players contribute $S_t$, and zero otherwise.

Players do not know the actions of others when they make their decisions. After each period, each player knows whether all members in his/her group (including himself/herself) have contributed in that period.

**Beliefs**

In contrast to the one-shot stage game, we consider a multi-period interaction between players where they can update their subjective beliefs regarding the behavior of others over time through observing the outcomes of the stage game in previous periods. In a typical player $i$'s mind, player $j$'s strategy type ($j \neq i$) is characterized by the highest level of contribution player $j$ (unconditionally) chooses. This belief structure on others' strategy type is a fundamental assumption of our belief-based learning model.

**Assumption 1 (Belief on Strategy Type).** In players' belief system, a player of strategy type $X$ will contribute $x$ for all $x \leq X$.[8]

When a player decides whether or not to contribute, what essentially matters is his/her belief about all other players' strategy types being no less than the stake level in the current period: when other players' strategy types are all greater than or equal to the stake level, they will surely contribute; when at least some player's strategy type is less than the stake level, such a player will not contribute.

Thus, in a period $t$, a player $i$'s belief on a contribution threshold $X$ can be characterized by his/her probability assessment that all other players' strategy type will be at least $X$, denoted by $G_t^i(X)$, where $G_t^i(\cdot)$ is a weakly decreasing function with $G_t^i(0)=1$ and $\lim_{X \to +\infty} G_t^i(X)=0$.[9]

---

[8] Crawford (1995) and Weber (2005, 2006) adopt a similar modeling methodology in coordination games assuming that players' discrete action is determined by rounding the continuous latent strategy variable that represents their beliefs. In our binary setting with varying stakes across periods, the strategy type X is such a latent strategy variable, where player $i$ with strategy type X is believed to contribute the stake $S_t$ in period $t$ if and only if $S_t \leq X$.

[9] In the earlier version of this paper, a level-k thinking model (Nagel, 1995; Stahl & Wilson, 1995; Ho et al., 1998; Costa-Gomes et al., 2001; Costa-Gomes & Crawford, 2006; Costa-Gomes et al., 2009) is assumed in order to derive the theoretical predictions of the model. We thank the editors and the referees for suggesting the direction for a simpler belief-updating learning process.

Relying on different combinations of game outcomes and the decision of *i*, including case 1 where everyone contributes, case 2 where *i* contributes while someone else does not, and case 3 where *i* does not contribute, his/her belief $G_t^i(\cdot)$ will have various updating paths. The details are relegated to the "Belief Updating" subsection in S3 of the Supplementary Material for interested readers.

**Mechanisms**

We consider three mechanisms that vary only in terms of the stake levels $S_t (t=1,\cdots,N)$ over periods. Under the **Big-Bang mechanism**, the coordination game starts at a high-stake level $\bar{S}$ and continues with the same stake level over time. Under the **Semi-Gradualism mechanism**, the coordination game starts at a low-stake level $\underline{S}$ but jumps to the high-stake level $\bar{S}$ after $N_1$ periods, where $N_1 < N$. Under the **Gradualism mechanism**, the coordination game starts at the low-stake level $\underline{S}$ and gradually increases the stake level over time at a constant rate such that after $N_1$ periods the stake level reaches $\bar{S}$ and stays at $\bar{S}$ afterwards.

**Results**

We now present the theoretical predictions of our belief-based model, relegating all the technical analyses and proofs to the "Main Results and Proofs" subsection of S3 in the Supplementary Material. Assuming that every player forms a belief regarding the minimum threshold that all other players will contribute, where the threshold refers to the strategy type described in Assumption 1, a player's optimal behavior critically depends on his/her beliefs and can be characterized in the following lemma.

**Lemma 1 (Optimal Behavior).** For a given period *t* with stake level $S_t$, player *i* chooses *C* if and only if he/she believes the probability of all other players' strategy type being at least $S_t$ is no less than $\alpha^{-1}$ (namely, $G_t^i(S_t) \geq \alpha^{-1}$). Thus, in the symmetric belief case where every player has the same belief ($G_t^i(S_t)=G_t(S_t)$), the equilibrium outcome is that every player contributes if $G_t(S_t) \geq \alpha^{-1}$ and no player contributes if $G_t(S_t) < \alpha^{-1}$.

The intuition of Lemma 1 is very simple: the more confident a player is about his/her group members contributing at the current stake level, the more likely this player is going to contribute, and the more likely the coordination will succeed.

We now investigate how different conditions on stake levels across periods will affect the likelihood of successful coordination. First, we show that coordination is easier when the stake size is small.

**Proposition 1 (Initial Stake).** The lower the stake at period 1, $S_1$, the higher the probabilities that each player will contribute and that the coordination will succeed at period 1.

To see why the above proposition holds, note that the weak monotonicity property of the belief function $G_t^i(\cdot)$ implies that a player is more confident about his/her group member contributing at a lower stake level than at a high stake level. This observation together with the insight from Lemma 1 provides the intuition behind Proposition 1.

Next we show that when fixing the size of stakes across different periods, the dynamic pattern of coordination is path-dependent and persistent.

**Proposition 2 (Persistency of Success/Failure).** In two consecutive periods with the same stake $S_{t+1} = S_t$, a group will succeed in coordination at period $t+1$ when it succeeds at period $t$, and will fail in coordination at period $t+1$ when it fails at period $t$.

We can also show that a slow increase in stake is always not worse than a quick increase for maintaining successful coordination, as indicated in Proposition 3 below.

**Proposition 3 (Quick vs. Slow Increase).** Conditional on successful coordination at period $t$, for given $S_t$ (Case a) or given $S_{t+1}$ (Case b), the smaller the stake difference between period $t$ and $t+1$, $S_{t+1} - S_t (> 0)$, the higher the probability that the coordination at period $t+1$ will succeed.

Based on the results described above, we would like to theoretically compare the coordination success rate of the three mechanisms in question (namely the Big-Bang mechanism, Semi-Gradualism mechanism, and the Gradualism mechanism).

**Proposition 4 (Performance Comparison).** For any number of players ($I \geq 2$), for any multiplier ($\alpha > 1$), and for any weakly decreasing belief functions ($G_1^i(\cdot), i = 1, \cdots, I$):

(a) Gradualism outperforms Semi-Gradualism; (b) Semi-Gradualism outperforms Big-Bang,

where mechanism A outperforms mechanism B if for period $t = N_1 + 1, \cdots, N$, with stake level $\bar{S}$, A succeeds in coordination whenever B succeeds in coordination.

Proposition 4 provides a complete ranking of the coordination success rates among the three mechanisms of interest, and shows that the Gradualism mechanism is superior to the other two in terms of promoting successful coordination at the final high-stake periods.[10]

## 4.2 Main Hypotheses in Both Stages

Based on the theoretical predictions of Proposition 4, we would like to directly test how these mechanisms perform in the laboratory environment, which is formally stated in the following hypothesis.

---

[10] We also conducted comparative statics analysis for the impacts of the number of players, the multiplier size, and the type of belief functions on the coordination success rate. We were also able to show that given strictly increasing belief functions, there always exists a gradualism mechanism where the stake (not necessarily evenly) increases overtime such that the success of coordination is achieved with certainty. All these results are included in the "Additional Results and Proofs" subsection of S3 in the Supplementary Material.

**Hypothesis 1: In Stage 1, the Gradualism treatment outperforms the Semi-Gradualism treatment and the Semi-Gradualism treatment outperforms the Big Bang treatments in the high-stake projects.**

When subjects enter the second stage, they may have learned about group members via coordination outcomes in the first stage and formed their beliefs about the general population regarding their contribution tendencies. The coordination performance in the first stage of the game can influence subjects' perception of new group members and subsequently how they play in the second stage.[11] Given that the Gradualism treatment may promote more successful group coordination (relative to other treatments) in the first stage, we propose that, conditional on being placed in the Gradualism treatment during the first stage, players will be more likely to contribute (relative to players from other treatments) when they enter the second stage of the game. Alternatively, players may not necessarily form their beliefs about new group members based on their coordination history with old members, however, reinforcement learning (e.g., Roth & Erev, 1995) can induce players with a coordination success (or failure) at the end of Stage 1 to (or not to) contribute when they enter the first period of Stage 2. Thus, we have the second hypothesis below:

**Hypothesis 2: Subjects who were in the Gradualism treatment in Stage 1 are more likely to contribute in the first period of Stage 2 than those subjects in other treatments in Stage 1.**

## 5. Results: Impact of Gradualism on Coordination

This section presents our results of coordination outcomes. We begin by focusing our analysis on the following three outcome variables per period: (1) whether a group coordinates successfully (defined as whether all four group members contribute) or not, (2) whether an individual contributes or not, and (3) the payoff of each individual.

### 5.1 First Stage Result Highlights

To test Hypothesis 1, we examine the performance of various treatments in Periods 7–12 of the first stage, when all treatments face the same high stake, which is the main interest of this study.

**Main Result: The Gradualism treatment significantly outperforms alternative treatments: starting at a low stake and growing slowly leads to more successful coordination and higher earnings in the high-stake periods.**

Figure 2 shows the success rate by treatment. In Period 7, 66.7% of Gradualism groups coordinate successfully (i.e., all four group members contribute), whereas the success rates of Big Bang, Semi-

---

[11] Other studies provide evidence that history can influence subsequent behavior. In a two-stage trust game, Bohnet and Huck (2004) find that once players get to experience a cooperative environment in the first stage of a game, they become more trusting (of others) in a new environment in the second stage.

Gradualism, and High Show-up Fee groups are only 16.7%, 33.3%, and 30%, respectively. Figure 2 illustrates that the success rates for all the treatments remain stable from Period 7 to Period 12. We use the average success rate over Periods 7–12 for each group as one observation, so each group has one observation. The average success rate over Periods 7–12 in the Gradualism treatment is higher than that in the Big Bang, Semi-Gradualism, and High Show-up Fee treatments (Wilcoxon-Mann-Whitney two-sided test: p<0.01, p=0.06, and p=0.09, respectively; observations are at the group level given that coordination success is a group-level outcome; N=64). Success rate in the Big Bang treatment is insignificantly different from that in the Semi-Gradualism and High Show-up Fee treatments (Wilcoxon-Mann-Whitney two-sided test: p=0.18, p=0.42, respectively). [12]

[Figure 2 about here]

Figure 3 shows the average individual earnings by treatment. Subjects in the Big Bang and High Show-up Fee groups have higher earning potential (i.e., higher stakes) in Periods 1–6. However, on average, they earn less than the subjects in the Gradualism treatment because of the high success rates in the Gradualism treatment. The Semi-Gradualism groups earn less than the Gradualism treatment in Periods 2–6 because of the lower earnings potential. These earning differences persist over Periods 7–12, when all the treatment groups experience the same highest stake. The differences in accumulative individual earnings over Periods 7–12 between Gradualism and Big Bang, Semi-Gradualism and High Show-up Fee treatments are all highly statistically significant (Wilcoxon-Mann-Whitney two-sided test for Period 7: p<0.0001, p<0.0001, and p=0.02, respectively, when observations are at the individual level, N=256; p-values become 0.05, 0.04 and 0.38, respectively, when observations are at the group level, N=64).[13] Clearly, from the perspective of social welfare, gradualism also works best.

[Figure 3 about here]

There may be a concern that the differences in performance over Periods 7–12 are caused by the effect of individual wealth (earnings) accumulation over Periods 1–6.[14] Section S4 in the Supplementary Material addresses this concern in detail: by both comparing the performance between the High Show-up Fee treatment and the other treatments and running regressions controlling for accumulated earnings, we show that the aforementioned Main Result is not driven by the wealth effect.

---

[12] It is noted that there is no statistically significant difference in the successful coordination rate between the Semi-Gradualism treatment and the two Big-Bang treatments, while Proposition 4 implies the former outperforms the latter. It is admitted that our theoretical comparison result is mainly qualitative instead of quantitative, and the size of difference in performance between these two treatments crucially depends players' belief functions at the initial stake. Since the initial stake in our experiment is quite small, it is reasonable to expect $G_1^i(S_1)$ to be close to 1, which leads to a small difference in performance between these two treatments. Please see the comment at the end of the proof of Proposition 4 in the S3 section in the Supplementary Materials for technical details.

[13] Earnings in the Big Bang treatment are insignificantly different from those in the Semi-Gradualism and High Show-up Fee treatments (Wilcoxon-Mann-Whitney two-sided test: p=0.72 and p=0.37, respectively).

[14] Such accumulated earnings may capture the impact of coordination history and play an important role in the coordination dynamics, thus its elimination may be unnecessary. However, if accumulated earnings have a direct effect beyond the history channel, then we should treat it cautiously.

Since the wealth effect does not drive our main result, we can combine the Big Bang and High Show-up Fee treatments and compare them together with the Gradualism treatment, leading to even higher statistical power (Wilcoxon-Mann-Whitney two-sided test regarding average success rate over Periods 7–12: $p<0.0001$ for individual-level, $p<0.01$ for group-level; Wilcoxon-Mann-Whitney two-sided test regarding accumulated individual earnings over Periods 7–12: $p<0.0001$ for individual-level, $p=0.07$ for group-level).

### 5.2  Coordination Dynamics in the First Stage

To identify why the Gradualism treatment performs best in Periods 7–12 relative to other treatments, we examine the coordination dynamics in Figures 2 and 4 (and Figures S1-1 to S1-6 as well as Figure S2 in the Supplementary Material). The three observations presented below support Propositions 1–3 in the theoretical model, respectively.

**Observation 1: The lower the stake size, the higher the average contribution and success rates in Period 1.**

Figure 4 displays contribution rates for Period 1. The average contribution rate is above 90% for Semi-Gradualism and Gradualism treatments with a low stake, which is much higher than the contribution rate of 60% for Big Bang and High Show-up Fee treatments with a high stake (Wilcoxon-Mann-Whitney two-sided test between these two categories: $p<0.0001$; observations are at the individual level).

[Figure 4 about here]

As shown in Figure 2, the differences in success rates are more evident. Over two-thirds of Semi-Gradualism and Gradualism groups coordinate successfully at the low initial stake, whereas only 16.6% (or 30%) of the Big Bang (or High Show-up Fee) groups succeed at the high initial stake (Wilcoxon-Mann-Whitney two-sided test between these two categories: $p<0.0001$; observations are at the group level). A weakest-link structure requires that all four group members contribute at the same time to make coordination a success, thus a difference in contribution rate may result in an even larger difference in success rate.[15]

---

[15] Assuming the probability of contributing is independent across members in a group (which is plausible in Period 1 because players are randomly assigned to groups and have not yet interacted with each other), the success rate should be the biquadrate of the contribution rate. As long as the contribution rates are sufficiently high, the difference in the success rate exceeds that in the contribution rate.

**Observation 2: Conditional on having failed coordination in Period *t*, most groups fail at the same or a higher stake in Period *t*+1; conditional on successfully coordinating in Period *t*, most groups succeed at the same stake in Period *t*+1.**

Figures S1-1 and S1-2 in the Supplementary Material present the coordination success rate in each period conditional on coordination failure and success in the previous period, respectively. The success rate is always smaller than 20% (and often equal to 0%) if a group fails in the previous period. Thus, once a group fails to coordinate, it rarely becomes successful thereafter. This pattern is likely due to players obtaining limited information feedback regarding the group outcome in each period: each member does not know how many group members contribute if the group fails in coordination. The above findings are consistent with the coordination literature, which largely concludes that a group which reaches an inefficient outcome is not likely to subsequently achieve efficiency with limited information feedback and without further mechanisms.

Similarly, once a group succeeds, it almost always remains successful at the same stake. The success rate is generally larger than 80% (and often equal to 100%) if a group succeeds in the previous period (except Period 7 for the Semi-Gradualism, when the stake jumps from 2 to 14). Figure S2 in the Supplementary Material provides further details of the coordination results for each group.

**Observation 3: Conditional on successfully coordinating in Period *t*, most groups succeed at a slightly higher stake in Period *t*+1. However, fewer groups remain successful at a much higher stake in Period *t*+1.**

We note a large gap in success rates between the Gradualism and the Semi-Gradualism treatments in Period 7, when the stake jumps from 2 to 14 for the latter treatment. Both treatments exhibit high success rates of approximately 70% for the first six periods. However, the success rate of the Semi-Gradualism treatment falls dramatically to only 33.3% in Period 7, whereas that of the Gradualism treatment remains at a higher level of 66.7% (Wilcoxon-Mann-Whitney two-sided test between these two treatments in Period 7: $p<0.05$; observations are at the group level).

Figure S1-2 in the Supplementary Material shows that conditional on coordination success in the previous period, the coordination success rate is always above 90% for Gradualism, whereas it is below 50% for Semi-Gradualism in Period 7 when the stake jumps. Table S4 in the Supplementary Material reports the formal regression results and shows that the performance difference across treatments is largely due to different coordination success rates in the previous period: contribution (success) rate is on average 71.4 (93.3) percentage points smaller if the group failed in coordination in the previous period, compared to the case when the group succeeded.

The results of Semi-Gradualism and Gradualism treatments in Periods 6 and 7 confirm the importance of increasing the stake slowly and avoiding shocks to the stake size. In fact, in Period 7 the contribution rate of the Gradualism treatment is only about five percentage points higher than that of the Semi-Gradualism treatment (Figure 4). Why does the success rate of the Semi-Gradualism treatment drop sharply from Period 6 to Period 7, whereas that of the Gradualism treatment remains high? Notably, a high contribution rate does not guarantee a high success rate because the latter requires that most groups have all four group members contributing at the same time. For the Semi-Gradualism treatment, a moderate 15 percentage points decrease in the contribution rate from Period 6 to Period 7 translates to a sharp 40 percentage points drop in the success rate, suggesting that a large portion of subjects who give up contributing in Period 7 come from previously successful groups: an unanticipated big jump in the stake causes some of the subjects in previously successful groups to be unwilling to continue contributing. Previously successful coordination established at low-stake periods is sabotaged even if only one of the four group members stops contributing. Figure S2 in the Supplementary Material further confirms this. Thus, avoiding shocks can help overcome coordination difficulty and bolster coordination success, which is exactly why the Gradualism treatment outperforms the Semi-Gradualism treatment.

Driven by Observations 2 and 3, the success rates are quite stable over all periods except for a drop from Period 6 to Period 7 for the Semi-Gradualism treatment. Conditional on successfully coordinating in Period 1, Big Bang, Gradualism, and High Show-up Fee groups usually remain successful in subsequent periods. Groups in the Big Bang and High Show-up Fee treatments exhibit lower success rates in Period 1 than groups in the Gradualism treatment and therefore, on average, perform worse than groups in Gradualism treatment at a high-stake level.

## 5.3 Second Stage Result Highlights

In this section, we examine whether the treatment type in the first stage influences individual behavior and outcomes in the second stage, when subjects are placed in a new group.

**Observation 4: Subjects exposed to the Gradualism treatment in Stage 1 are more likely to contribute upon entering a new group in Stage 2 than subjects who are previously exposed to other treatments. However, this difference quickly disappears.**

Figure 4 shows that the subjects in the Gradualism treatment are 12.2 percentage points (86.1 vs. 73.9; $p<0.05$) more likely to contribute in the first period of Stage 2. However, the higher contribution rate at the beginning of Stage 2 of those Gradualism subjects translates into neither a higher success rate (Figure 2) nor higher average earnings (Figure 3). On average, Gradualism subjects earn 1.5 points less than subjects from other treatments (the difference is statistically insignificant.) As a direct consequence, the contribution rate of Gradualism subjects decreases faster than that of other subjects from Period 1 to Period 2 of Stage 2 and becomes comparable to that of other subjects in Period 2 and throughout Stage 2 (Figure 4). This convergence of behavior suggests a belief updating process that, after observing the coordination outcome in Period 1 of Stage 2, the subjects from the Gradualism treatment update their beliefs, perceiving their new group partners less likely to contribute than their first stage partners; therefore, they become less willing to contribute in subsequent periods.

Section S6 in the Supplementary Material provides more results regarding the second stage.

## 6. Belief Elicitation Experiment: Evidence of Belief-based Learning

In order to further explore whether the belief-based learning model outlined in Section 4 does indeed provide the underlying reason why gradualism works, we conducted a supplementary experiment at Xiamen University in May 2013, in which we adopted an identical design as the main experiment except that we elicited the belief of subjects about the action of others in each period.[16]

We have 24 subjects in the elicitation experiment (two groups each for Big Bang, Semi-Gradualism, and Gradualism treatments). These subjects are different from those in the main experiment. After each period, we ask the belief of each subject about the number of contributors among the other three group members, $N_C$, in his/her group (note that the group size is four). Section S7 in the Supplementary Material reports the detailed belief elicitation process.

We check whether the decisions of the subjects are consistent with their beliefs in each period. Then, we examine whether they have updated their beliefs based on the outcome feedback after each period. We present the results in Tables 3 and 4, respectively. Considering that estimating the treatment effect is not the purpose of the supplementary experiment, we pool all subjects and all periods in two stages and gain a total of 480 subject-period observations.

Table 2 shows how belief affects the contribution decision of the subjects in that period. We report OLS results for easier interpretation. As a robustness check, we also conduct Firth logit regressions (Heinze & Schemper, 2002) because the contribution variable is binary and the logit regressions encounter the problem of separation (Heinze & Schemper, 2002), which leads to qualitatively similar results (see Table S6 in the Supplementary Material). All five specifications with different control variables consistently show that the contribution decisions of the subjects in each period are indeed largely affected by their beliefs that all the other three group members contribute in that period.

Table 3 shows the belief formation and updating process. Column 1 examines the initial belief that all the other three group members contribute in Period 1. A higher stake in Period 1 leads to a smaller belief that all other three group members contribute, although the coefficient is statistically insignificant (note that we only have 24 observations). Columns 2 through 5 examine the belief updating process with different control variables. All results show that beliefs indeed are updated based on the previous belief and the coordination outcome.

---

[16] We ended up not eliciting beliefs in the main experiment because we were concerned that belief elicitation would have contaminated our results. Such a concern is verified by Rutström and Wilcox (2009) and Gächter and Renner (2010).

To conclude, the results provide suggestive evidence for the belief-based learning model. Of course, the purpose of our model was to illustrate how gradualism might work, but not to present a model of the actual belief-formation process taking place in the experiment. In addition, our evidence is only suggestive because we cannot rule out the possibilities that the actual decision is independent of beliefs, and that the beliefs are ex-post rationalizations when they are elicited.

## 7. Discussion and Concluding Remarks

In this study, we investigate the effects of gradualism – defined as increasing, step-by-step, the stake level required for group coordination projects – on successful group coordination using data from a randomized laboratory experiment. No previous study has identified which pattern of successive and exogenously set stake levels produces successful group coordination. We find strong evidence that gradualism can serve as a powerful mechanism in achieving socially optimal outcomes in group coordination. Gradualism significantly outperforms alternative paths, which shows that starting at a low-stake level and slowly growing the stake size are both important for coordination at later periods and at a substantial high-stake level.

We propose a belief-based learning model as one underlying channel that explains why gradualism leads to successful group coordination at a substantial high-stake level. The results from a supplementary experiment with belief elicitation support this explanation. With that said, although the model provides a useful framework for rationalizing the patterns in the data, we cannot definitively rule out alternative channels that can also explain our main result.

The two key features of the gradualism mechanism in the current study are a) the binary choice set in each period, and b) the gradual increase in the stake of projects. First, the subjects are restricted to two choices in each period (without mandatory or semi-mandatory institutions, such as sanctions and social pressures): to contribute the exact stake size or not to contribute at all. The binary choice set prevents players from contributing too little to harm others and from contributing too much to get hurt, thus serving as an effective coordination device. Second, a small initial stake leads to a high belief that others will contribute and thus encourages players to contribute at the beginning, whereas a gradual path of increasing stake maintains such an optimistic belief as well as the high willingness of the players to contribute even when the stake becomes substantial.

Our second finding relates to the spillover effect of coordination experience across social groups. Subjects treated in the gradualism setting are more likely to contribute upon entering a new group than subjects under different treatments. However, when these gradualism subjects find that their contributions have not been rewarded in the new groups, they reduce their tendency to contribute. Thus, the contribution behavior of the subjects treated in the gradualism regime and alternative regimes converges quickly. This result may have policy implications for ensuring efficient coordination outcomes when society members from diverse cultures and institutions merge.

Our findings have broad implications for how managers can structure team assignments optimally before the team performs crucial high-stake tasks. Specifically, our central finding regarding gradualism underscores the role that supervisors and managers can play in leading teams to successfully coordinate in high-stake tasks and achieve higher productivity: teams should start in a low-stake situation and then slowly move to tasks involving higher stakes. Our study demonstrates that gradualism can be a valuable tool for aiding high-stake coordination: a leader can start a group with an easier version of a task to provide reassurance and strengthen beliefs.

Various management practices, from finance to law enforcement to venture capital, already exhibit this stylized gradualist feature of group coordination. In industries that significantly rely on effective coordination – such as consulting, information technology, engineering, or medical care (Gittell, 2002; Faraj & Xiao, 2006) – our findings point to a promising approach that teams can use so as to gradually work up to larger projects with bigger efforts and investments. In real world teams, gradualism may be even more effective because it would also enable managers to swap out low-performing employees or disband dysfunctional teams if they fail to perform in low-stake situations before the stakes become too high. Therefore, managers can employ the gradualism feature as a risk-mitigation strategy. Although there are other methods that managers can use to increase productivity in the organizations they oversee, our study provides one feasible and effective method that enhances the design of group training, improves overall team effort and increases subsequent team productivity.

Charness, G. (2000). Self-serving cheap talk: A test of Aumann's conjecture. *Games and Economic Behavior, 33(2),* 177-194.

Chen, R., & Chen, Y. (2011). The potential of social identity for equilibrium selection. *American Economic Review*, *101 (6),* 2562-2589.

Chen, Y., Li, S. X., Liu, T. X., & Shih, M. (2014). Which hat to wear? Impact of natural identities on coordination and cooperation. *Games and Economic Behavior*, *84*, 58-86.

Choi, S., Gale, D., & Kariv, S. (2008). Sequential equilibrium in monotone games: A theory-based analysis of experimental data. *Journal of Economic Theory, 143(1),* 302-330.

Cooper, R., DeJong, D. V., Forsythe, R., & Ross, T. W. (1992). Communication in coordination games. *Quarterly Journal of Economics, 107(2),* 739–771.

Costa-Gomes, M. A., Crawford, V. P., & Broseta, B. (2001). Cognition and behavior in normal-form games: An experimental study. *Econometrica, 69(5),* 1193-1235.

Costa-Gomes, M. A., & Crawford, V. P. (2006). Cognition and behavior in two-person guessing games: An experimental study. *American Economic Review, 96(5),* 1737-1768.

Costa-Gomes, M. A., Crawford, V. P., & Iriberri, N. (2009). Comparing models of strategic thinking in Van Huyck, Battalio, and Beil's coordination games. *Journal of European Economic Association*, *7*(2-3), 365-376.

Crawford, V. P. (1995). Adaptive dynamics in coordination games. *Econometrica, 63(1),* 103-143.

Crawford, V. P., & Iriberri, N. (2007). Level-k auctions: Can a non-equilibrium model of strategic thinking explain the winner's curse and overbidding in private-value auctions? *Econometrica*, *75(6),* 1721-1770.

Devetag, G. (2003). Coordination and information in critical mass games: An experimental study. *Experimental Economics, 6*(1), 53–73.

Devetag, G. (2005). Precedent transfer in coordination games: An experiment. *Economics Letters, 89*(2), 227-232.

Devetag, G., & Ortmann, A. (2007). When and why? A critical survey on coordination failure in the laboratory. *Experimental Economics*, *10*(3), 331-344.

Dorsey, R. (1992). The voluntary contributions mechanism with real time revisions. *Public Choice, 73(3),* 261-282.

Duffy, J., & Feltovich, N. (2002). Do actions speak louder than words? Observation vs. cheap talk as coordination devices. *Games and Economic Behavior, 39(1),* 1–27.

Duffy, J., & Feltovich, N. (2006). Words, deeds, and lies: Strategic behavior in games with multiple signals. *Review of Economic Studies*, *73*(3), 669-688.
23

**Figure 1: Stake Patterns of the Treatments**

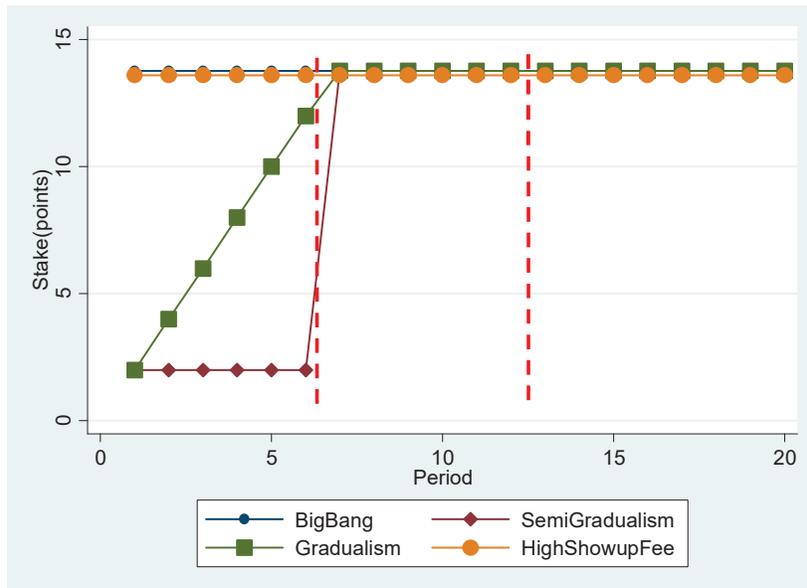

Note: The first and second vertical dotted lines separate the two halves (Periods 1–6 and 7–12) of the first stage, and the two stages (Periods 1–12 and 13–20), respectively. Coordination performance of different treatments in the second half (Periods 7-12) of the first stage is the main interest of this study. The High Show-up Fee treatment is identical to the Big Bang treatment except for a higher show-up fee (see Table 1 for details). Group members are fixed within each stage, whereas subjects from various treatments are reshuffled when they enter the second stage.

**Figure 2: Success Rates of Groups by Treatment and Period**

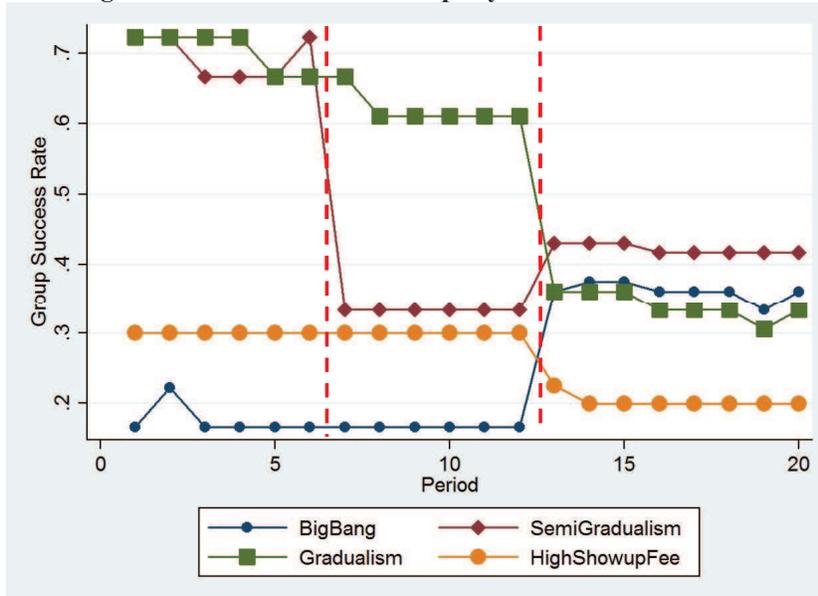

Note: A group coordinates successfully if all four members contribute the stake in that period.

**Figure 3: Average Individual Earning by Treatment and Period**

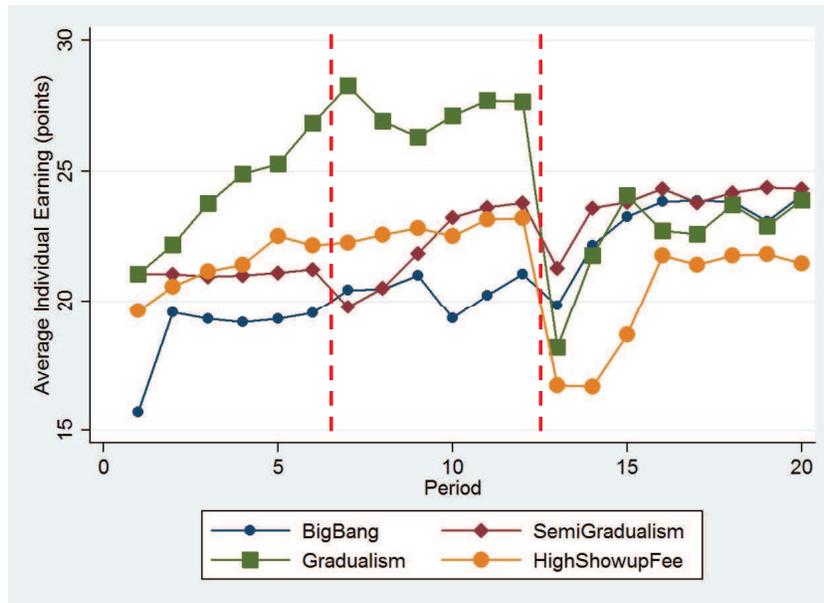

**Figure 4: Contribution Rate by Treatment and Period**

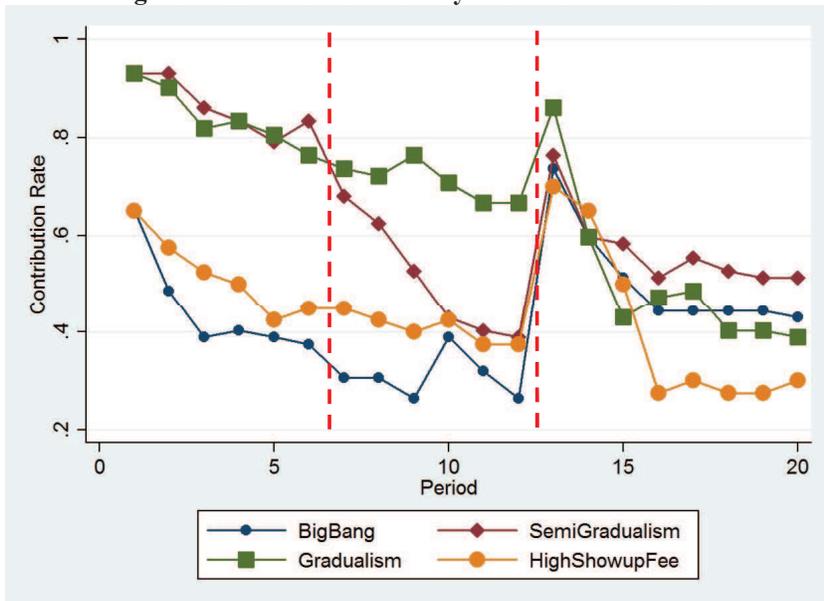

Table 1: Summary of Treatments in the First Stage

| Treatment | Big Bang | Semi-Gradualism | Gradualism | High Show-up Fee |
|---|---|---|---|---|
| Endowment in each period (points) | 20 | 20 | 20 | 20 |
| Show up Fee (points) | 400 | 400 | 400 | 480 |
| Exchange Rate (points/CNY) | 40 | 40 | 40 | 40 |
| Stake in Period 1 (points) | 14 | 2 | 2 | 14 |
| Stake in Period 6 (points) | 14 | 2 | 12 | 14 |
| Stake in Period 7-12 (points) | 14 | 14 | 14 | 14 |
| Number of groups | 18 | 18 | 18 | 10 |
| Number of subjects | 72 | 72 | 72 | 40 |
| Average earning up to Period 6 (points; excluding show-up fee) | 112.42 | 126.31 | 143.94 | 127.35 |
| Median earning up to Period 6 (points; excluding show-up fee) | 106 | 130 | 162 | 106 |

Note: We have 18 sessions in total. Ten sessions have 16 subjects for each session who are randomized into the above four treatments, and the other eight sessions have 12 subjects for each session who are randomized into the Big Bang, Semi-Gradualism, and Gradualism treatments.

**Table 2: The Effect of Belief on Contribution Decisions in Belief Elicitation Sessions (OLS)**

| | Contribution Dummy | | | | |
|---|---|---|---|---|---|
| | (1) | (2) | (3) | (4) | (5) |
| Belief that all other three contribute | 0.916*** | 0.847*** | 0.841*** | 0.723*** | 0.518*** |
| | (0.050) | (0.069) | (0.067) | (0.082) | (0.106) |
| Stake | -0.012** | -0.011** | 0.001 | 0.001 | 0.006 |
| | (0.005) | (0.005) | (0.007) | (0.007) | (0.006) |
| Subject fixed effects | N | Y | Y | Y | Y |
| Period fixed effects | N | N | Y | Y | Y |
| Lagged contribution dummy | | | | 0.213** | 0.106 |
| | | | | (0.096) | (0.080) |
| Lagged success dummy | | | | | 0.318** |
| | | | | | (0.116) |
| Constant | 0.291*** | 0.0871 | 0.131* | -0.0704 | -0.0755 |
| | (0.0729) | (0.0698) | (0.0758) | (0.0893) | (0.0817) |
| Observations | 480 | 480 | 480 | 456 | 456 |
| R-squared | 0.669 | 0.731 | 0.762 | 0.786 | 0.801 |

Note: Robust standard errors in parentheses are clustered at the individual level. *** $p<0.01$, ** $p<0.05$, * $p<0.1$. Contribution is a dummy indicating whether the subject contributes in the current period. Belief ranges from 0 to 1 and represents the subject's belief that all other three group members contribute in current period. Stake refers to the stake in current period. Firth logit regressions (Heinze & Schemper, 2002) in Table S6 of the Supplementary Material report qualitatively similar results.

**Table 3: Belief Formation and Updating in Belief Elicitation Sessions (OLS)**

| | Belief that all other three group members contribute in current period | | | | |
|---|---|---|---|---|---|
| | Period 1 | Period 2-20 (Both Stages Together) | | | |
| | (1) | (2) | (3) | (4) | (5) |
| Lagged belief | | 0.372*** | 0.373*** | 0.260*** | 0.269*** |
| | | (0.070) | (0.0683) | (0.061) | (0.057) |
| Lagged success dummy | | 0.490*** | 0.488*** | 0.566*** | 0.561*** |
| | | (0.064) | (0.0623) | (0.066) | (0.062) |
| Stake | -0.011 | | | | |
| | (0.014) | | | | |
| ∆Stake | | | 0.002 | 0.001 | 0.000 |
| | | | (0.015) | (0.015) | (0.016) |
| Subject fixed effects | N | N | N | Y | Y |
| Period fixed effects | N | N | N | N | Y |
| Constant | 0.693*** | 0.061*** | 0.060*** | 0.102*** | 0.143*** |
| | (0.094) | (0.016) | (0.014) | (0.008) | (0.042) |
| Observations | 24 | 456 | 456 | 456 | 456 |
| R-squared | 0.032 | 0.777 | 0.777 | 0.808 | 0.818 |

Note: Robust standard errors in parentheses are clustered at the individual level. *** $p<0.01$, ** $p<0.05$, * $p<0.1$. Belief ranges from 0 to 1. Stake refers to the stake in current period. ∆Stake is the difference in stake between current period and the previous period.